\documentclass[fleqn,10pt]{wlscirep}
\title{Broken selection rule in the quantum Rabi model}

\author[1,2,*]{P. Forn-D\'iaz}
\author[3]{G. Romero}
\author[2]{C. J. P. M. Harmans}
\author[4,5]{E. Solano}
\author[2]{J. E. Mooij}
\affil[1]{Institute for Quantum Computing, Department of Physics and Astronomy, and Waterloo Institute for Nanotechnology, University of Waterloo, Waterloo, N2L 3G1, Canada}
\affil[2]{Kavli Institute of Nanoscience, Delft University of Technology, Post Office Box 5046, 2600 GA Delft, The Netherlands}
\affil[3]{Departamento de F\'isica, Universidad de Santiago de Chile (USACH), Avenida Ecuador 3493, 9170124, Santiago, Chile}
\affil[4]{Department of Physical Chemistry, University of the Basque Country UPV/EHU, Apartado 644, E-48080 Bilbao, Spain}
\affil[5]{IKERBASQUE, Basque Foundation for Science, Mar\'ia D\'iaz de Haro 3, E-48013 Bilbao, Spain}
\affil[*]{pforndiaz@uwaterloo.ca}

\begin{abstract}
Understanding the interaction between light and matter is very relevant for fundamental studies of quantum electrodynamics and for the development of quantum technologies. The quantum Rabi model captures the physics of a single atom interacting with a single photon at all regimes of coupling strength. We report the spectroscopic observation of a resonant transition that breaks a selection rule in the quantum Rabi model, implemented using an $LC$ resonator and an artificial atom, a superconducting qubit. The eigenstates of the system consist of a superposition of bare qubit-resonator states with a relative sign. When the qubit-resonator coupling strength is negligible compared to their own frequencies, the matrix element between excited eigenstates of different sign is very small in presence of a resonator drive, establishing a sign-preserving selection rule. Here, our qubit-resonator system operates in the ultrastrong coupling regime, where the coupling strength is 10\% of the resonator frequency, allowing sign-changing transitions to be activated and, therefore, detected. This work shows that sign-changing transitions are an unambiguous, distinctive signature of systems operating in the ultrastrong coupling regime of the quantum Rabi model. These results pave the way to further studies of sign-preserving selection rules in multiqubit and multiphoton models.
\end{abstract}
\begin{document}

\flushbottom
\maketitle

\thispagestyle{empty}

\section*{Introduction}

The quantum Rabi model describes the dipolar interaction between a two-level system, a qubit of frequency $\omega_q$, and a quantized electromagnetic field mode, an oscillator of frequency $\omega_r$ \cite{braak, kike_viewpoint}. The strength of the interaction $g$ between qubit and oscillator defines different regimes of coupling, reflecting the dominating energy scales in the dynamics of the system. The limit of coupling strength $g$ much smaller than qubit/resonator frequencies $g\ll\lbrace\omega_r,\omega_q\rbrace$ is the well-known Jaynes-Cummings (JC) model \cite{JC}, where the rotating-wave approximation (RWA) holds. The ultrastrong coupling (USC) regime is defined by $g/\omega_r\ge0.1$. Such condition implies that counter-rotating terms in the interaction Hamiltonian cannot be neglected. The USC regime presents interesting features such as a non-trivial ground state with photonic excitations \cite{ciuti_prb}, the generation of qubit-oscillator nonclassical states \cite{sahel} and ultrafast gates for quantum computation \cite{guille_ultrafast}. The intricate ground-state structure has motivated various proposals for quantum simulation \cite{tureci, egger}. Despite its fundamental interest and applications, the experimental study of the USC regime is challenging. In this sense, superconducting quantum circuits \cite{blais} are suitable systems to achieve large coupling strengths, despite some intrinsic limitations for most superconducting qubit types \cite{devoret}. In fact, recent ideas have emerged \cite{bourassa, borja_prl, borja_usc} on how to reach the USC regime using superconducting flux qubits. The latter can attain a large galvanic coupling to resonators \cite{bloch-siegert} and have a huge anharmonicity, typically larger than $20~\rm{GHz}\gg\omega_q/2\pi$, both of which are crucial to reach $g/\omega_r>0.1$. USC in an open system has recently been reported using a superconducting flux qubit \cite{USC-TL}.

In the JC model $g/\omega_r\ll1$, the energy-level structure consists of a manifold of dressed-state doublets each containing $n$ excitations $|n,\pm\rangle$ \cite{JC}. Each doublet $|n\rangle$ is labeled with a quantum number \mbox{$\pm$} corresponding to the relative sign in the superposition of uncoupled qubit-resonator states $\{|e,n-1\rangle,|g,n\rangle\}$:
\begin{eqnarray}
\label{ds1}|n,+\rangle_{\rm{JC}} &=& \cos(\theta_n/2)|e, n-1\rangle + \sin(\theta_n/2)|g,n\rangle,\\
\label{ds2}|n,-\rangle_{\rm{JC}} &=& \sin(\theta_n/2)|e, n-1\rangle - \cos(\theta_n/2)|g,n\rangle.
\end{eqnarray}
The mixing angle is defined as \mbox{$\tan\theta_n=2g\sqrt{n}/(\omega_q-\omega_r)$}. In the case of resonator driving represented by the operator $H_{\rm{d}}\sim(a+a^{\dag})$, transitions between dressed states of different sign, $|n,\pm\rangle_{\rm{JC}}\leftrightarrow|n\pm1,\mp\rangle_{\rm{JC}}$, are much less favored than those preserving it, $|n,\pm\rangle_{\rm{JC}}\leftrightarrow|n\pm1,\pm\rangle_{\rm{JC}}$. For USC rates, \mbox{$g/\omega_r\sim0.1$}, one can still effectively describe the spectrum of the system using dressed-state doublets $|\widetilde{n,\pm}\rangle$ with correspondingly larger transition matrix elements and corrections of order $g/(\omega_r+\omega_q)$ to manifolds with total number of excitations $n\pm2$ \cite{klimov, BS_Grifoni, dissipation_ciuti, dissipation_Blais}. The considerations on sign-changing transitions in the USC regime do not apply. Thus for coupling strength much smaller than qubit/resonator frequencies, one can establish a selection rule for the case of driven resonator $H_{\rm{d}}\sim (a + a^{\dag})$. In the USC regime, due to the different nature of the system eigenstates, the selection rule may be broken. For qubit driving $H_{\rm{d}}\sim\sigma_x$, the selection rules are slightly different than the resonator driving case, as detailed in the results section. 

Observations of sign-preserving transitions have been reported for superconducting transmon qubits coupled to transmission line resonators with $g/\omega_r\sim0.02$ \cite{fink_ladder, fink_thermal}. To date, only a few experimental systems have reached the USC regime. In superconducting circuits, artificial atoms have been ultrastrongly coupled to resonators \cite{niem, bloch-siegert, baust, sembausc} and transmission lines \cite{USC-TL}, where new physics beyond the RWA were reported. Polaritons in quantum wells have been ultrastrongly coupled to microcavities, causing large frequency shifts due to enhanced nonlinearities \cite{anappara}. Finally, electrons in a two-dimensional electron gas undergoing cyclotron transitions $\omega_{\rm c}$ were coupled to metamaterial cavities where the coupling strength achieved was $g/\omega_{\rm c}\sim0.58$ \cite{faist}. None of the above cited experiments reported observations or was able to provide evidence of sign-changing transitions.

On the theory side, despite the considerable amount of recent progress on studying the physics of the USC regime \cite{jorge_dsc, braak, ridolfo_thermal, tureci, sahel_superradiance, daniel_prx, crespi_rabi, dereli}, only recently transitions between dressed states have been carefully analyzed \cite{simone_parity}.

Here, we report the spectroscopy of a qubit-resonator system in the USC regime of the quantum Rabi model. We use the fact that our system is at finite temperature to excite several transitions between dressed states. One of the transitions is shown to break the sign-preserving selection rule, demonstrating another paradigmatic example of the unique physics occurring in the USC regime. Using a superconducting flux qubit coupled to a lumped-element $LC$ resonator we are able to reproduce the quantum Rabi model of a qubit coupled to a single oscillator \cite{braak}. In our experimental system, the qubit and the resonator have no higher energy levels within the relevant frequency range, which is advantageous for observing the breakdown of the sign-selection rule.

\section*{Results}

\subsection*{Theoretical model}
Our experiment operates in the regime of $g/\omega_r\sim0.1$, which will be referred to as the quantum Bloch-Siegert (QBS) regime. The eigenstates of the system in this regime are analogous to the doublet structure of dressed states of the JC model (see the methods section for more details):
\begin{eqnarray}
\label{qbs1}
|\widetilde{n,+}\rangle = \cos(\varphi_n/2)\left(|e, n-1\rangle+\lambda\sqrt{n-1}|g,n-2\rangle\right) + \sin(\varphi_n/2)\left(|g,n\rangle-\lambda\sqrt{n+1}|e,n+1\rangle\right),\\
\label{qbs2}
|\widetilde{n,-}\rangle = \sin(\varphi_n/2)\left(|e, n-1\rangle + \lambda\sqrt{n-1}|g,n-2\rangle\right) - \cos(\varphi_n/2)\left(|g,n\rangle-\lambda\sqrt{n+1}|e,n+1\rangle\right).
\end{eqnarray}
With respect to the JC model, a correction of order $\lambda\equiv g/(\omega_r+\omega_q)$ appears connecting to states containing $n\pm2$ number of excitations due to the effect of the counter-rotating terms. The mixing angle is here defined as
\begin{equation}
\tan\varphi_n = \frac{2g_n\sqrt{n}}{\delta_n},
\end{equation}
which is different from the JC in a photon-number dependent detuning $\delta_n \equiv \delta + 2n\omega_{\mathrm{BS}}$ and a photon-number dependent coupling strength $g_n\equiv\sqrt{n}g-n^{3/2}g\omega_{\rm{BS}}/(\omega_r+\omega_q)$. $\delta \equiv\omega_q-\omega_r$ is the detuning in the JC model, $\omega_{\rm{BS}}\equiv g^2/(\omega_r+\omega_q)$ is the Bloch-Siegert shift that originates from the counter-rotating terms. The existence of a selection rule for transitions between dressed states of different sign depends on the relative magnitude of matrix elements for sign-changing/preserving transitions. The matrix elements $T_{Xi\leftrightarrow j}=\langle\widetilde{i}|H_d|\widetilde{j}\rangle$ between dressed states for resonator-type driving $H_d/\hbar = A_r(a+a^{\dag})\cos(\omega_d t)$ can be calculated to be
\begin{eqnarray}
\label{Tel_r}
T_{X|\widetilde{n,+}\rangle\leftrightarrow|\widetilde{n+1,+}\rangle}\sim\cos(\varphi_n/2)\cos(\varphi_{n+1}/2)\sqrt{n}\left(1+\lambda^2(n-1)\right)+\sin(\varphi_n/2)\sin(\varphi_{n+1}/2)\sqrt{n+1}\left(1+\lambda^2(n+2)\right)\nonumber\\-\lambda\sin(\varphi_n/2)\cos(\varphi_{n+1}/2),\\
T_{X|\widetilde{n,-}\rangle\leftrightarrow|\widetilde{n+1,-}\rangle}\sim\sin(\varphi_n/2)\sin(\varphi_{n+1}/2)\sqrt{n}\left(1+\lambda^2(n-1)\right)+\cos(\varphi_n/2)\cos(\varphi_{n+1}/2)\sqrt{n+1}\left(1-\lambda^2(n+2)\right)\nonumber\\+\lambda\sin(\varphi_{n+1}/2)\cos(\varphi_n/2),\\
T_{X|\widetilde{n,+}\rangle\leftrightarrow|\widetilde{n+1,-}\rangle}\sim\cos(\varphi_n/2)\sin(\varphi_{n+1}/2)\sqrt{n}\left(1+\lambda^2(n-1)\right)-\sin(\varphi_n/2)\cos(\varphi_{n+1}/2)\sqrt{n+1}\left(1+\lambda^2(n+2)\right)\nonumber\\-\lambda\sin(\varphi_{n+1}/2)\sin(\varphi_n/2),\\
\label{trpm}
T_{X|\widetilde{n,-}\rangle\leftrightarrow|\widetilde{n+1,+}\rangle}\sim\sin(\varphi_n/2)\cos(\varphi_{n+1}/2)\sqrt{n}\left(1+\lambda^2(n-1)\right)-\cos(\varphi_n/2)\sin(\varphi_{n+1}/2)\sqrt{n+1}\left(1+\lambda^2(n+2)\right)\nonumber\\+\lambda\cos(\varphi_{n+1}/2)\cos(\varphi_n/2).
\end{eqnarray}
Clearly, the sign-preserving transitions are always possible while the sign-changing transitions are less likely to be excited. For vanishing qubit-resonator coupling $g$ and $\delta>0$ ($\delta<0$), $\varphi_n\to0$ ($\varphi_n\to\pi$), so $\sin(\varphi_n/2)\to0$ ($\sin(\varphi_n/2)\to1$), $\cos(\varphi_n/2)\to1$ ($\cos(\varphi_n/2)\to0$). This implies that sign-changing transitions disappear when the coupling strength is much smaller than qubit/resonator frequencies, establishing a selection rule for this type of transitions.

For single qubit driving $H_d/\hbar = A_q\sigma_x\cos(\omega_d t)$ the transition matrix elements $T_{\sigma_xi\leftrightarrow j}$ between dressed states read
\begin{eqnarray}
T_{\sigma_x|\widetilde{n,+}\rangle\leftrightarrow|\widetilde{n+1,+}\rangle}&\sim&+\sin(\varphi_n/2)\cos(\varphi_{n+1}/2)+\lambda(\sqrt{n}\cos(\varphi_n/2)\cos(\varphi_{n+1}/2)-\sqrt{n+1}\sin(\varphi_n/2)\sin(\varphi_{n+1}/2)),\\
T_{\sigma_x|\widetilde{n,-}\rangle\leftrightarrow|\widetilde{n+1,-}\rangle}&\sim&-\cos(\varphi_n/2)\sin(\varphi_{n+1}/2)+\lambda(\sqrt{n}\sin(\varphi_n/2)\sin(\varphi_{n+1}/2)-\sqrt{n+1}\cos(\varphi_n/2)\cos(\varphi_{n+1}/2)),\\
T_{\sigma_x|\widetilde{n,+}\rangle\leftrightarrow|\widetilde{n+1,-}\rangle}&\sim&+\sin(\varphi_n/2)\sin(\varphi_{n+1}/2)+\lambda(\sqrt{n}\cos(\varphi_n/2)\sin(\varphi_{n+1}/2)+\sqrt{n+1}\sin(\varphi_n/2)\cos(\varphi_{n+1}/2)),\\
\label{Tel_q} T_{\sigma_x|\widetilde{n,-}\rangle\leftrightarrow|\widetilde{n+1,+}\rangle}&\sim&-\cos(\varphi_n/2)\cos(\varphi_{n+1}/2)+\lambda(\sqrt{n}\sin(\varphi_n/2)\cos(\varphi_{n+1}/2)+\sqrt{n+1}\cos(\varphi_n/2)\sin(\varphi_{n+1}/2)).
\end{eqnarray}
For $\delta<0$, as is the case in our experiment, $T_{\sigma_x|\widetilde{n,-}\rangle\leftrightarrow|\widetilde{n+1,+}\rangle}/T_{\sigma_x|\widetilde{n,-}\rangle\leftrightarrow|\widetilde{n+1,-}\rangle}\sim1/\tan(\varphi_{n+1}/2)\to0$ as $g\to0$, implying a sign-preserving rule. However, the other transitions do not follow the same rule since $T_{\sigma_x|\widetilde{n,+}\rangle\leftrightarrow|\widetilde{n+1,-}\rangle}/T_{\sigma_x|\widetilde{n,+}\rangle\leftrightarrow|\widetilde{n+1,+}\rangle}\sim\tan(\varphi_{n+1}/2)\to\infty$ as $g\to0$. Clearly for single-qubit driving, the selection rule considerations of driven resonator do not apply. In the rest of this work we only discuss the transitions $|1,-\rangle\leftrightarrow|2,-\rangle$ and $|1,-\rangle\leftrightarrow|2,+\rangle$ that follow the sign-preserving rule both for resonator-driven as well as qubit-driven transitions. The case of single-qubit $\sigma_z$ driving is described in the Methods. Therefore all our conclusions on broken selection rules still hold.
Throughout the rest of the text we will drop the tilde to refer to the dressed states in the USC regime simply as $|n,\pm\rangle$.

Analyzing spectra in the USC regime requires accounting for the out-of-equilibrium quantum dynamics featuring the coloured nature of dissipative baths. One possibility is to project the master equation in the dressed-state basis $|n,\pm\rangle$ \cite{dissipation_Blais}. Another possibility is the second-order time-convolutionless projection operator method (TCPOM)~\cite{QOSBook}, which has been proven useful to correctly describe the open system dynamics in the USC regime~\cite{dissipation_ciuti,nataf_ciuti}. Here, we follow the latter approach where the master equation reads
\begin{equation}
\frac{d\hat{\rho}}{dt}=\frac{1}{i\hbar}[\hat{H}_0+\hat{H}_{\rm drive},\hat{\rho}]+\sum_{r=a,x,z}\hat{U}_r\hat{\rho} \hat{S}_r+\hat{S}_r\hat{\rho} \hat{U}_r^{\dag}-\hat{S}_r\hat{U}_r\hat{\rho}-\hat{\rho} \hat{U}_r^{\dag}\hat{S}_r,
\label{MasterEq}
\end{equation}    
with $\hat{S}_a=\hat{a}+\hat{a}^{\dag}$, $\hat{S}_x=\hat{\sigma}_x$, $\hat{S}_z=\hat{\sigma}_z$, and the operators $\hat{U}_k$ are defined as
\begin{equation}
\hat{U}_k  =\int^{\infty}_{0}\nu_k(\tau)e^{(-i/\hbar)\hat{H}_0\tau}\hat{S}_ke^{(i/\hbar)\hat{H}_0\tau}d\tau,\\
\nu_k(\tau) =\int^{\infty}_{-\infty}\Gamma_k(\omega)\{n_k(\omega)e^{i\omega\tau}+[n_k(\omega)+1]e^{-i\omega\tau}\}d\omega.
\label{Eqnu}
\end{equation}  
We consider thermal baths at temperature $T$ for all dissipative channels, where the relaxation coefficients $\Gamma_k(\omega)=2\pi d_k(\omega)\alpha^2_k(\omega)$ depend on the spectral density of the baths $d_k(\omega)$ and the system-bath coupling strength $\alpha_k(\omega)$. Moreover, the system Hamiltonian includes the qubit-resonator interaction as well as the presence of classical microwave fields, that is   
\begin{equation}
\label{rabi_ham}\hat{H}_0=\frac{\hbar\omega_q}{2}\hat{\sigma}_z+\hbar\omega_r \hat{a}^{\dag}\hat{a} +\hbar g(\cos\theta_q\sigma_z-\sin\theta_q\sigma_x)(\hat{a}+\hat{a}^{\dag}),
\end{equation}
\begin{equation} H_{\rm d}=\hbar A_{\rm qb}\cos(\omega_Dt)\hat{\sigma}_x+\hbar A_r\cos(\omega_Dt)(\hat{a}+\hat{a}^{\dag}).
\end{equation}
The qubit mixing angle $\theta_q$ is defined as $\tan\theta_q = \Delta/\epsilon$, with $\Delta$ being the tunnel coupling between the qubit eigenstates and $\hbar\epsilon=2I_p(\Phi-\Phi_0/2)$ the magnetic energy of the qubit (see Methods). The out-of-equilibrium dynamics assumes a perturbative treatment of driving mechanisms which is valid for amplitudes $\{A_{\rm qb},A_r\}\ll\omega_q,\omega_r,g$~\cite{blockade_ridolfo}, as is the case throughout this work. In all numerical simulations we use up to 6 Fock states in the resonator basis which is sufficient for $g/\omega_r\sim0.1$.

\subsection*{Sign-preserving transition}\label{sec3}

In order to explore dressed-state transitions, we use a superconducting flux qubit galvanically coupled to an $LC$ resonator. The results presented here are obtained with the same device where the Bloch-Siegert was first reported \cite{bloch-siegert}. All relevant experimental details can be found there. The resonator has frequency $\omega_r/2\pi = 8.13$~GHz, while the qubit has persistent current $I_p=500$~nA and a minimal splitting of $\Delta/2\pi = 4.2$~GHz, so the detuning is $\delta=\Delta - \omega_r = -2\pi\times3.93$~GHz at the symmetry point of the qubit. The coupling strength is $g/2\pi = 0.82$~GHz. The latter implies $g/\omega_r\simeq0.1$, so the system operates in the USC regime. The qubit and the resonator are thermally anchored to the mixing chamber stage of a dilution refrigerator at 20~mK. The qubit state is read out using a DC-SQUID magnetometer. An external superconducting coil is used to set the magnetic flux in the qubit loop near $\Phi = \Phi_0/2$. At this flux bias point the effects of the counter-rotating terms of the Hamiltonian are strongest (equation~(16) with $\sin\theta_q = 1$).

A numerical diagonalization of the Hamiltonian in equation~(16) leads to the spectrum shown in Fig.~1(a). 
The solid lines correspond to transitions from the ground state, while the dashed lines are transitions from the first excited state. The intermediate dashed transition $|1,-\rangle\leftrightarrow|2,-\rangle$ (see Fig.~1(b)) around 8~GHz is the one studied in detail in this section and corresponds to a sign-preserving transition. The highest-energy, dashed transition $|1,-\rangle\leftrightarrow|2,+\rangle$ is studied in the next section and is the first transition in ascending energy to break the sign-preserving selection rule, given our choice of parameters. At the qubit symmetry point $\Phi=\Phi_0/2$ only transitions connecting states of different parity are allowed \cite{blais_pra}. Therefore, the red sideband $|1,-\rangle\leftrightarrow|1,+\rangle$ and blue sideband $|0,g\rangle\leftrightarrow|2,-\rangle$ are forbidden at that point. Outside the symmetry point all transitions are permitted due to the asymmetry in the flux qubit potential \cite{nori_prl, deppe_natphys}. At the qubit symmetry point $\Phi=\Phi_0/2$, the system eigenstates correspond to those in Eqs.~(3),~(4) and the selection rules introduced in the previous section are therefore expected.
\begin{figure}
\centering
\includegraphics{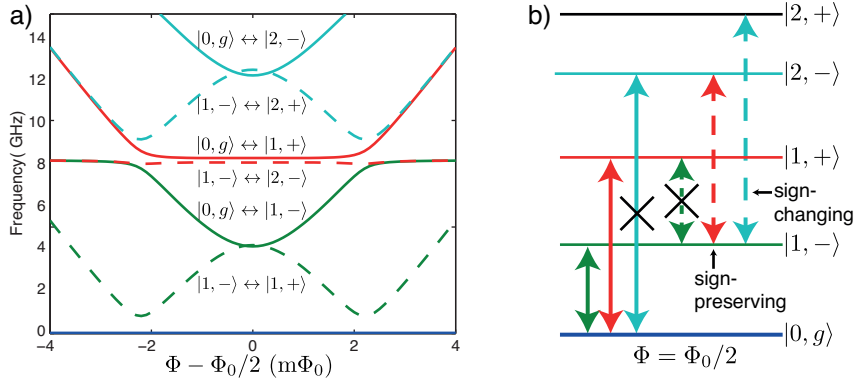}
\caption{Qubit spectrum and transitions. (a) Numerical diagonalization of system Hamiltonian in equation~(16), as function of external magnetic flux. Solid lines correspond to transitions from the ground state. Dashed lines correspond to transitions from the first excited state. (b) Energy levels at the symmetry point, with the sign-changing and sign-preserving transitions highlighted. $|1,-\rangle\leftrightarrow|1,+\rangle$ and $|0,g\rangle\leftrightarrow|2,-\rangle$ are forbidden here by parity selection rules.}
\end{figure}

Fig.~2(a) shows the spectrum of the system at low power near the resonator frequency $\omega_r/2\pi = 8.13$~GHz. The solid and dashed curves are a fit of equation~(16). The dotted line is the JC model with the same fitted parameters. The difference between the JC and Rabi models is the Bloch-Siegert shift $\omega_{\rm{BS}}$ \cite{bloch-siegert}, with a maximum of $55~$MHz at $\Phi=\Phi_0/2$. A negative (positive) shift of the frequency corresponds to a resonator (qubit)-like dressed state transition. Fig.~2(b) shows a spectroscopy trace at $\Phi = \Phi_0/2$. The measurements are performed following a similar protocol as the one developed previously for a tunable-gap flux qubit \cite{arkady} (see also Methods). Following from Fig.~1, we find that the higher frequency resonance in Fig.~2(b) corresponds to the transition $|0,g\rangle\leftrightarrow|1,+\rangle$, which given the detuning $\delta/2\pi = -3.93$~GHz is mostly a resonator-like excited dressed state. The lower frequency resonance corresponds to $|1,-\rangle\leftrightarrow|2,-\rangle$, therefore it is a sign-preserving transition. The finite temperature in our system enables the excitation of this transition. 
The signal from this resonance is weak given the small amount of population of qubit-like state $|1,-\rangle$ in thermal equilibrium. The fact that we can observe this transition at low driving strength indicates a large matrix element as predicted by equations~(6-13). The low signal-to-noise ratio leads to spurious peaks and dips which do not represent additional transitions, as verified by the spectrum in Fig.~1. The geometry of our system is such that qubit and resonator driving amplitudes have a comparable value, $A_{\rm{qb}}\simeq A_r$ (see Fig.~6 in the Methods), therefore both systems are simultaneously driven in all measurements presented.
\begin{figure}[!hbt]
\centering
\includegraphics{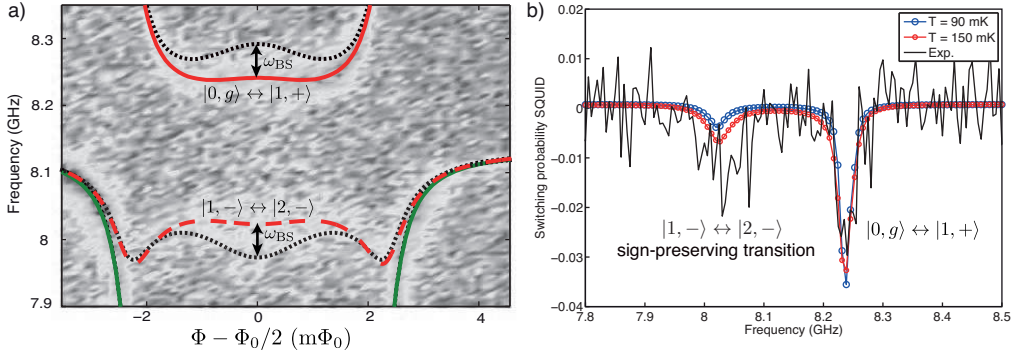}
\caption{Sign-preserving transition. (a) Measured spectrum with fit of equation~(16). Solid lines correspond to transitions from the ground state, cf.~Fig.~1. The dotted lines correspond to the JC model using the fitted parameters from equation~(16). The red-dashed line is the sign-preserving transition $|1,-\rangle\leftrightarrow|2,-\rangle$. Near $\Phi=\Phi_0/2$ the deviation from the JC model is given by the Bloch-Siegert shift $\omega_{\rm{BS}}\equiv g^2/(\omega_r+\omega_q)\simeq2\pi\times55~$MHz. (b) Spectrum at $\Phi=\Phi_0/2$. The switching probability of the SQUID detector is referenced to 50~\%. A numerical simulation of master equation equation~(14) is shown for $T=90~$mK ($T=150~$mK) with blue asterisks (red circles). The low amplitude of the sign-preserving transition at 8.02~GHz is related to the qubit thermal population in equilibrium.}
\end{figure}

The resonances in this spectrum look broad due to the low relaxation time of this particular device, mostly due parasitic capacitance to the SQUID readout circuitry. Other flux qubit experiments have shown much better performance, even with SQUID detectors \cite{jonas, simon_rot}. There is a clear asymmetry between the two resonances in Fig.~2(b). This difference allows us to approximately calibrate the electronic temperature of the system. Solving the master equation iteratively to reach steady-state is a lengthy process, making it impractical to run a fit of the observed spectrum. Instead, the unknown parameters are swept in order to find the optimal set that leads to a calculated spectrum closest to the measurements. We plot the expectation value of the qubit current operator $\langle \hat{I}_{\mathrm{circ}}\rangle\sim\mathrm{Tr}(\hat{\rho}\hat{\sigma}_z)$, where $\hat{\rho}$ is the steady-state solution of the TCPOM master equation, equation~(14) using the eigenstates from equation~(16). More details can be found in the Methods. The optimal set of parameters is $\Gamma_r/2\pi\simeq1$~MHz, $\Gamma_1/2\pi\simeq15$~MHz, $A_{\mathrm{qb}}/2\pi=A_r/2\pi\simeq12$~MHz. For the temperature, we plot $T=90$~mK (blue asterisks) and $T=150$~mK (red circles) to indicate the possible range of temperatures. Other experiments using superconducting qubits have observed comparable finite electronic temperatures \cite{fink_q-to-c}.

The results of the simulation clearly indicate an asymmetry between the two resonances near 8~GHz. A global scaling factor and an offset have been applied to the numerical solution to match the amplitude of the experimental resonances. The scaling factor gives the transduction between average persistent current in the qubit to SQUID switching probability. In the experiment, the offset value is chosen to be 50\% probability of the DC-SQUID to switch to the running state where a finite voltage is generated \cite{simon_rot}. The differences between simulation and experiment for the resonance at 8.02~GHz is most likely due to the low signal-to-noise, leading to an imprecise determination of the temperature of the system.

\subsection*{Broken sign-preserving selection rule}\label{sec4}
In this section, we perform spectroscopy of the qubit-resonator system with driving amplitude approximately 5 times larger than the one used in Fig.~2. The enhanced driving strength enables additional transitions to occur, both one- and two-photon transitions. The high-power spectrum of the system is shown in Fig.~3(a),
showing a rich spectral structure. For frequencies below 8~GHz, multi-photon processes can be identified. The frequency of each multi-photon transition follows the proper scaling $\omega_{n-}/n=\omega_r-g\sin\theta_q/\sqrt{2}$, as was already reported for a transmon qubit coupled to a resonator \cite{bishop_nlrabi}. $\theta_q$ is the qubit mixing angle defined below equation~(16). The resonance at 8.25~GHz from Fig.~2 has been power-broadened, overlapping with the sign-preserving transition $|1,-\rangle\leftrightarrow|2,-\rangle$ at 8.02~GHz. The large signal generated by this resonance may be related to a buildup of the population in the resonator due to the strong drive  \cite{reed_fidelity}. 
\begin{figure}[!hbt]
\centering
\includegraphics{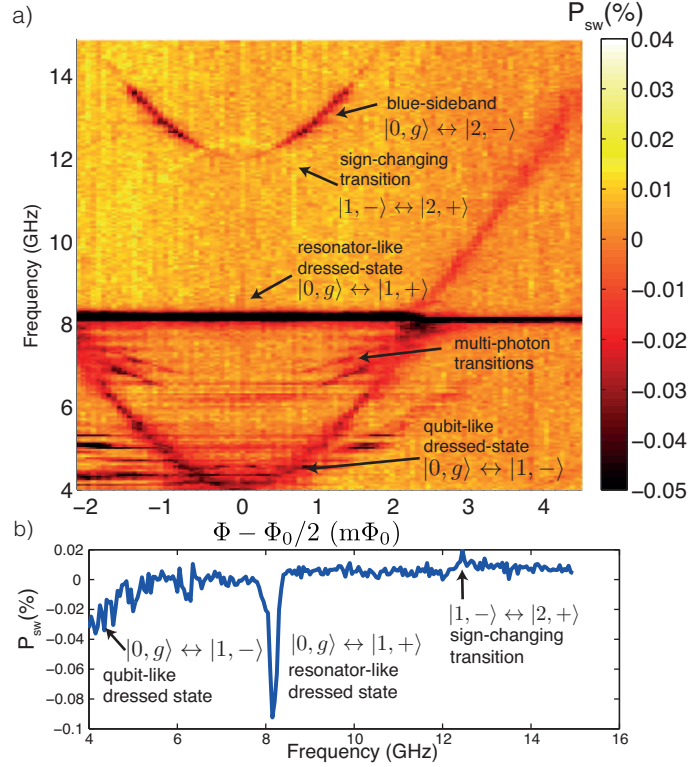}
\caption{Broken sign-preserving selection rule. (a) Spectrum of the qubit-resonator system at high driving power. Multi-photon transitions to higher-photon number dressed states are visible below the resonance at 8~GHz. Above 12~GHz, the blue sideband $|0,g\rangle\leftrightarrow|2,-\rangle$ can be clearly observed, together with an additional weaker resonance. The latter corresponds to $|1,-\rangle\leftrightarrow|2,+\rangle$, breaking the sign-preserving selection rule. The transition disappears away from the symmetry point due to the decrease of the coupling strength with the mixing angle in equation~(16), and the increase of dephasing rate of the qubit \cite{jonas, JL_dec, FQ_saclay}. (b) Spectroscopy trace at the symmetry point of (a). Above 12~GHz, the transition $|1,-\rangle\leftrightarrow|2,+\rangle$ is visible as a peak in switching probability.}
\end{figure}

Above 8~GHz, there are two distinctive resonances near $\Phi\simeq\Phi_0/2$. The transition increasing in frequency away from the symmetry point corresponds to the blue sideband, $|0,g\rangle\leftrightarrow|2,-\rangle$. The reason it vanishes at $\Phi=\Phi_0/2$ is due to parity selection rules \cite{blais_pra}. The other weaker resonance corresponds to the transition $|1,-\rangle\leftrightarrow|2,+\rangle$ and is, therefore, a sign-changing transition that breaks the sign-preserving selection rule explained in the introductory section. To our knowledge, this is the first observation of this kind of transitions between excited states of the quantum Rabi model. The small signal from this resonance evidences the difference in magnitude of the transition matrix elements as compared to the one for $|1,-\rangle\leftrightarrow|2,-\rangle$ given the difference in driving amplitudes used to acquire the data in Fig.~2 and Fig.~3.

The spectroscopic sign of any resonance detected by the SQUID magnetometer relates to the magnetic field generated by the qubit in steady-state when an external drive is present. The transition $|1,-\rangle\leftrightarrow|2,+\rangle$ increases the switching probability above the reference at 50\%, compared to all other resonances where the switching probability decreases below 50\%. This implies that in steady-state the qubit is more polarized to the ground state than in thermal equilibrium, contrary to the other resonances where the qubit ends up with more population in the excited state. The resonance $|1,-\rangle\leftrightarrow|2,+\rangle$ therefore cools the qubit down by transferring its excess thermal energy to the resonator, since the state $|2,+\rangle$ is a resonator-like excited dressed state. A spectroscopy trace at the symmetry point can be observed in Fig.~3(b). The switching probability at 12.3~GHz increases with respect to the background set at $P_{\mathrm{sw}}=0~\%$ (referenced to $50~\%$). 

\section*{Discussion}

In Figs.~4(a),(b) we calculate the matrix element for transitions $T_{|1-\rangle\leftrightarrow|2-\rangle}$, $T_{|1-\rangle\leftrightarrow|2+\rangle}$, respectively, as function of normalized coupling strength $g/\omega_r$ for driving the resonator $H_d\sim\hat{X}\equiv(a+a^{\dag})$. The blue solid line is calculated numerically from equation~(16), the green-dashed line corresponds to the quantum Bloch-Siegert (QBS) regime Eqs.~(7), (8), (11), (12) and the red-dotted line is the Jaynes-Cummings model calculated from the states in Eqs.~(1), (2). In Figs.~4(c), (d) we also plot the matrix elements $T_{|1-\rangle\leftrightarrow|2-\rangle}$, $T_{|1-\rangle\leftrightarrow|2+\rangle}$ for the case of single-qubit driving $H_d\sim\sigma_x$. The difference in matrix elements for sign changing/preserving transitions is very clear for both types of driving. Figs.~4(b), (d) show that in our experiment the broken-sign transition is mostly driven by the direct qubit drive. As shown in Fig.~4(e), the relative weight $T_{|1-\rangle\leftrightarrow|2-\rangle}/T_{|1-\rangle\leftrightarrow|2+\rangle}$ decreases with increasing $g/\omega_r$ for both types of driving since we consider finite detuning $\delta/2\pi = -3.91$~GHz. If we had the resonance condition $\omega_r = \omega_q=2\pi\times8.13~$GHz, the relative weight of the matrix elements for resonator (qubit) driving would reach $T_{|1,-\rangle\leftrightarrow|2,-\rangle}/T_{|1,-\rangle\leftrightarrow|2,+\rangle}\approx7 (1)$ at $g/\omega_r=0.1$, instead of 234(4.8) as shown in Fig.~4(e). The dashed vertical line in Figs.~4(a)-(e) marks the operating point of our device. The observation of the sign-changing transition $|1,-\rangle\leftrightarrow|2,+\rangle$ in this experiment can therefore be attributed to the system operating in the USC regime. 
\begin{figure}[!hbt]
\centering
\includegraphics{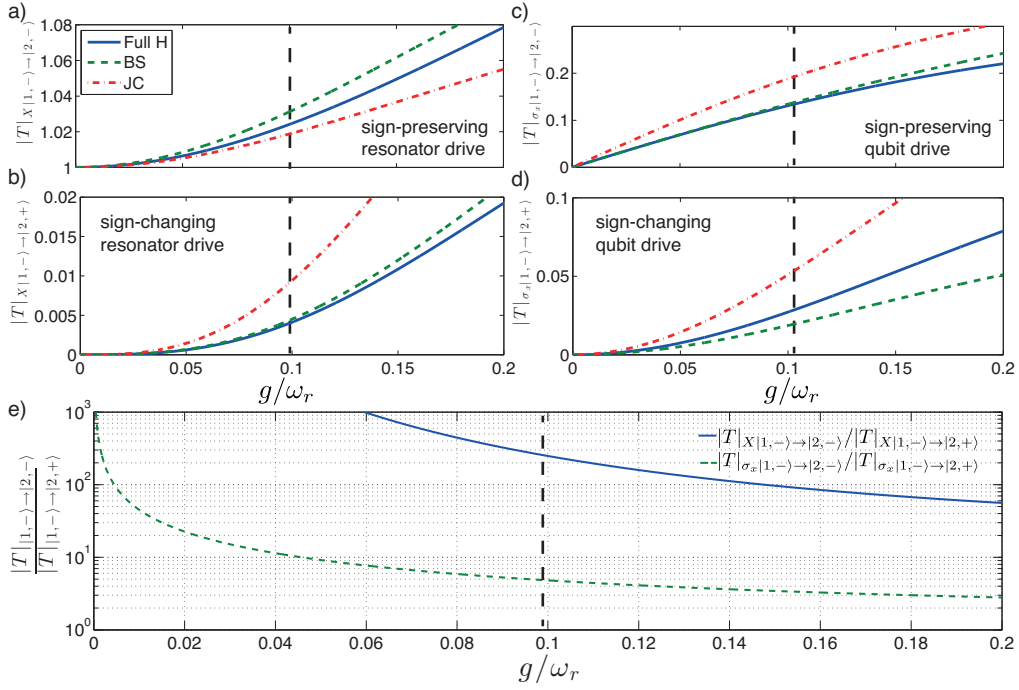}
\caption{Matrix elements calculation. Driven resonator transitions $|1,-\rangle\leftrightarrow|2,+\rangle$ (a) and $|1,-\rangle\leftrightarrow|2,-\rangle$ (b). (c), (d) correspond to the same transitions as in (a), (b) for the case of single-qubit driving. Equation~(16) is numerically solved (solid line) and compared to the Bloch-Siegert (BS) Hamiltonian Eqs.~(7), (8), (11), (12) (dashed line) and the Jaynes-Cummings (JC) Hamiltonian (dash-dotted line). (e) Relative weight $T_{|1-\rangle\leftrightarrow|2-\rangle}/T_{|1-\rangle\leftrightarrow|2+\rangle}$ of the matrix elements calculated in (a)-(d) for resonator (solid blue) and qubit (dashed green) driving. The value decays due to the increased similarity between the structure of the eigenstates when approaching the deep strong coupling regime \cite{braak, jorge_dsc}. The vertical dashed lines mark the operating conditions of the current experiment.}
\end{figure}

In Fig.~5 we compare the measured spectrum near 12~GHz (Fig.~5(a)) with a simulation (Fig.~5(b)) with the TCPOM master equation~(14) using larger drive amplitude, $A_{\mathrm{qb}}/2\pi = A_r/2\pi = 50$~MHz, keeping the rest of terms equal as those found in the previous section, with $T=100$~mK. The increased driving amplitude used here is approximately a factor of 5 larger compared to the amplitude used for the simulations in Fig.~2(b), closely matching the ratio of driving amplitudes used in the measurements of Fig.~3 relative to those in Fig.~2. The results are plotted in Fig.~5(b) for the probability to find the system excited, $1-|\langle0,g|\rho|0,g\rangle|^2$, where $\rho$ is the density matrix of the coupled system calculated with the master equation in steady-state. The position of the main two resonances appearing in the experiment in Fig.~5(a) are clearly reproduced up to $\sim-1~$m$\Phi_0$, particularly at the symmetry point $\Phi=\Phi_0/2$ where only the sign-changing transition $|1,-\rangle\leftrightarrow|2,+\rangle$ is excited. Beyond $-1~\rm{m}\Phi_0$, the experimental signal in Fig.~5(a) disappears due to an increase of the dephasing rate of the qubit away from the symmetry point \cite{jonas, JL_dec, FQ_saclay} which is not taken into account in the numerical simulations. 
\begin{figure}[!hbt]
\centering
\includegraphics{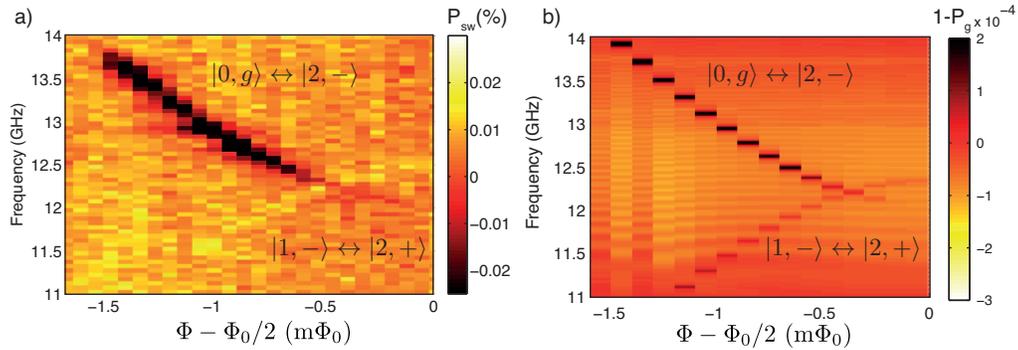}
\caption{Numerical simulation of qubit spectrum. (a) Zoom-in of the broken sign-preserving selection rule transition. (b) Numerical simulation using equation~(14). The $z$-scale corresponds to probability to find the system excited, $1-P_g =1- |\langle0,g|\rho|0,g\rangle|^2$, where $\rho$ is the steady-state solution of the master equation. The transition $|1,-\rangle\leftrightarrow|2,+\rangle$ breaking the sign-preserving selection rule is clearly visible at all fluxes while the blue sideband generates much more signal away from the symmetry point. For this simulation $T=100$~mK, $A_{\mathrm{qb}}/2\pi = A_r/2\pi = 50$~MHz.}
\end{figure}

\section*{Conclusions}

We report the observation of transitions between excited states of a superconducting flux qubit-resonator system in the ultrastrong coupling regime. The strength of the coupling combined with initial thermal population of the qubit permits the observation of a transition between excited states that was never detected before. Such transition changes the relative sign in the superposition of bare qubit-resonator states in the dressed-state level structure. We developed a theoretical model based on the time convolutionless projection operator method that reproduces all transitions observed spectroscopically. Our work, therefore, verifies the existence of sign-changing transitions in the ultrastrong coupling regime of the quantum Rabi model, despite their weak strength. Until now, evidence for superconducting qubit-resonator systems to operate in the USC regime was found as spectral deviations from the Jaynes-Cummings model \cite{bloch-siegert, niem, baust}. We instead put forward the observation of sign-changing transitions in the quantum Rabi model as a direct signature of any physical system in the USC regime for resonator-type driving. Our work can be extended to multi-qubit \cite{simone_parity} or multi-photon \cite{simone_collapse} quantum Rabi models where forbidden transitions play a key role in understanding and manipulating the energy-level structure of the system.
\section*{Methods}

\subsection*{Quantum Bloch-Siegert Hamiltonian and eigenenergies}
\label{appA}
For a qubit coupled to a resonator the Hamiltonian that describes the full system dynamics is the quantum Rabi model \cite{braak}. When $g/\omega_{r},g/\omega_q\ll1$ the rotating-wave approximation (RWA) holds and the Jaynes-Cummings (JC) model is obtained \cite{JC}. In the regime $g/\omega_r\sim g/\omega_q\sim0.1$ the RWA is no longer valid but a perturbative treatment still permits deriving the Hamiltonian of the system analytically \cite{klimov, BS_Grifoni}, as well as a model on dissipation \cite{dissipation_ciuti, dissipation_Blais, blockade_ridolfo}. 

The Bloch-Siegert Hamiltonian derives from the quantum Rabi Hamiltonian via perturbative treatment of the parameter $\lambda=g/(\omega_q + \omega_r)\ll1$ \cite{klimov,dissipation_Blais, BS_Grifoni}. In this perturbative regime, the counter-rotating terms are treated as off-resonant fields and the quantum Rabi model can be transformed in an effective dispersive picture. The resulting Hamiltonian is:
\begin{equation}
\label{app_22}\hat{H}_{\mathrm{BS}}/\hbar=-\frac{\omega_q}{2}\hat{\sigma}_z+\omega_r\left(\hat{a}^{\dag}\hat{a} + \frac{1}{2}\right) +\omega_{\mathrm{BS}}\left[-\hat{\sigma}_z\left(\hat{a}^{\dag}\hat{a} +\frac{1}{2}\right) - \frac{1}{2}\right] + g(\hat{n})(\hat{a}^{\dag}\hat{\sigma}^- + \hat{a}\hat{\sigma}^+),
\end{equation}
where the Bloch-Siegert shift $\omega_{\mathrm{BS}}$ is defined as 
\begin{equation}\omega_{\mathrm{BS}}=\frac{g^2\sin^2\theta_q}{\omega_q + \omega_r},\end{equation}
with photon-dependent coupling strength $g(\hat{n})$
\begin{equation}\label{g_BS}g(\hat{n}) = g\sin\theta_q\left(1-\hat{a}^{\dag}\hat{a}\frac{\omega_{\mathrm{BS}}}{\omega_q+\omega_r}\right).\end{equation}
Here $\omega_q=\sqrt{\Delta^2+\epsilon^2}$ is the flux qubit splitting with $\hbar\epsilon=2I_p(\Phi-\Phi_0/2)$ being the magnetic energy of the qubit in its truncated two-state Hilbert space. 
The qubit mixing angle $\theta_q$ is defined as $\tan\theta_q=\Delta/\epsilon$.

Projecting the Hamiltonian on the product state $|\Psi\rangle = |i,n\rangle,~i=\{e, g\}$ has a box-diagonal representation that simplifies the full diagonalization:
\begin{equation}
\left(
\begin{array}{c c c c c}
-\frac{\delta}{2} - \omega_{\mathrm{BS}} & 0  & \dots & & 0\\
0 & \left(\begin{array}{c c} \displaystyle\frac{\omega_q + \omega_r}{2} & g(1) \\  g(1) & -\displaystyle\frac{\delta}{2} + \omega_r - 2\omega_{\mathrm{BS}}\end{array}\right) & & \\
\vdots &   \hspace{5cm}\ddots & & &\vdots\\
0&\dots&  \left(\begin{array}{c c} \displaystyle\frac{\delta}{2}+n(\omega_r+\omega_{\mathrm{BS}}) - \omega_{\mathrm{BS}} & g(n)\sqrt{n} \\ g(n) \sqrt{n} & \displaystyle-\frac{\delta}{2}+n(\omega_r - \omega_{\mathrm{BS}}) - \omega_{\mathrm{BS}} \end{array}\right)&\dots&0\\
\vdots & \dots & &  \ddots&\vdots\\
 \end{array}\right).
\end{equation}
For box $n$, the following eigenvalues are found:
\begin{equation}\label{eqbslevels1}
E_{\pm,n}=n\omega_r - \omega_{\mathrm{BS}} \pm\frac{1}{2}\sqrt{\left(\delta + 2n\omega_{\mathrm{BS}}\right)^2 + 4(g\sin\theta_q)^2n\left(1 - n\frac{\omega_{\mathrm{BS}}}{\omega_q+\omega_r}\right)^2},\end{equation}
with ground state
\begin{equation}\label{eqbslevels2}
E_{g,0} = -\frac{\delta}{2}-\omega_{\mathrm{BS}},\end{equation}
where $n\ge1$ represents the number of total excitations in the uncoupled basis.
All levels are shifted by $\omega_{\mathrm{BS}}$ as compared to the JC model. In addition, both the effective detuning between levels $\delta + 2n\omega_{\mathrm{BS}}$ as well as the effective coupling constant (second term in the square root of equation~(22)) depend on the number of excitations. The latter is the second-order correction in this dispersive picture. Using the following definitions:
\begin{equation}\delta_n = \delta + 2n\omega_{\mathrm{BS}},\end{equation}
\begin{equation}g_n = g\sin\theta_q\sqrt{n}\left(1-n\frac{\omega_{\mathrm{BS}}}{\omega_q+\omega_r}\right),\end{equation}
\begin{equation}\omega_{r,n}=n\omega_r-\omega_{\mathrm{BS}},\end{equation}
the eigenenergies take the form
\begin{equation}E_{\pm, n}=\omega_{r, n}\pm\frac{1}{2}\sqrt{\delta_{n}^2+4g_n^2},\end{equation}
exactly analogous to the JC model \cite{blais}. Following this analogy the eigenstates of the system can be expressed similarly to the eigenstates of the JC model, with corrections to states with different total number of excitations, as
\begin{eqnarray}
\label{dressed_states1}|\widetilde{n,+}\rangle = \cos(\varphi_n/2)\left(|e, n-1\rangle+\lambda\sqrt{n-1}|g,n-2\rangle\right) + \sin(\varphi_n/2)\left(|g,n\rangle-\lambda\sqrt{n+1}|e,n+1\rangle\right),\\
\label{dressed_states2}|\widetilde{n,-}\rangle = \sin(\varphi_n/2)\left(|e, n-1\rangle + \lambda\sqrt{n-1}|g,n-2\rangle\right) - \cos(\varphi_n/2)\left(|g,n\rangle-\lambda\sqrt{n+1}|e,n+1\rangle\right).
\end{eqnarray}
where the mixing angle is defined as
\begin{equation}
\tan\varphi_n = \frac{2g_n\sqrt{n}}{\delta_n}.
\end{equation}
Equations~(28),~(29) show that the energy eigenstates of the Bloch-Siegert Hamiltonian correspond to doublets of superpositions exactly as with the Jaynes-Cummings model, with coefficients having a different dependence on the number of photons $n$ and corrections of order $\lambda\equiv g/(\omega_r+\omega_q)$ to other states with different number of excitations. Therefore the selection rule considerations between transitions that change the sign quantum number explained in the introduction break down in the USC regime.

\subsection*{Dissipation dynamics}
Here, the parameters of the qubit-resonator system described in the main text are derived using numerical simulations. Besides the temperature of the bath $T$, there are a set of unknown parameters that need to be determined: the qubit relaxation and decoherence rates $\Gamma_1, \Gamma_2$, the resonator decay rate $\Gamma_r$ and the amplitudes of the external fluxes $A_{\mathrm{qb}}$ and $A_r$ driving the qubit and the resonator, respectively. Estimates of the mutual inductance between qubit/resonator and microwave line permit setting $A_{\mathrm{qb}} = A_r$ by geometry (see Fig.~6). We also assume that qubit decoherence is governed by relaxation alone $\Gamma_2 = \Gamma_1/2$ since this device shows very short $T_1\sim10~$ns times near the symmetry point $\Phi=\Phi_0/2$.  
\begin{figure}[!hbt]
\centering
\includegraphics{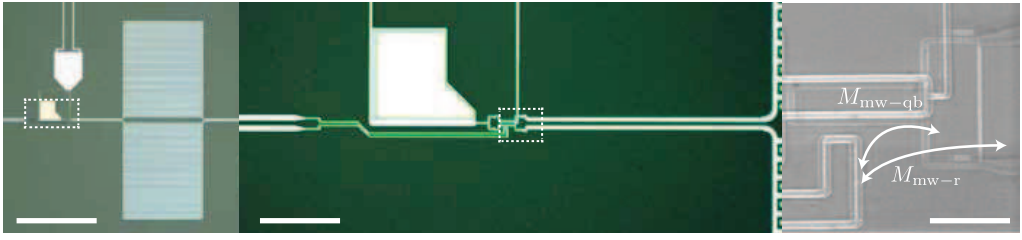}
\caption{Device images. Left and center images are optical pictures of the device. (Left) The large squares are the interdigitated finger capacitor of the resonator. (Center) The white block in the center is a gold heat sink for the SQUID magnetometer. Scale bar is 150~$\mu$m ($40~\mu$m) for the left (center) picture. (Right) Scanning electron microscope image of the qubit and the SQUID sitting on top of it. The microwave antenna is the loop seen at the bottom-left with mutual inductance to the qubit $M_{\rm{mw-qb}}$ and to the resonator $M_{\rm{mw-r}}$. Scale bar is 5~$\mu$m.}
\end{figure}

A technique to characterize the temperature of a transmon qubit by driving Rabi oscillations was already developed \cite{fink_q-to-c}, but due to the short relaxation time of our qubit this technique is not suitable. We would like to emphasize that the coherence of the flux qubit used in this experiment is limited by spontaneous emission to the readout circuitry of the SQUID detector \cite{simon_rot}. Using proper filtering \cite{jcup} and a more symmetric circuit design \cite{arkady} would allow better quantitative study of the spectroscopic resonances. Other experiments have shown much longer values of qubit lifetime at the symmetry point \cite{jonas, JL_dec, FQ_saclay}, exceeding 10~$\mu$s.

We run a numerical simulation of the master equation presented in the theory section and compare the steady-state solution to the spectroscopy measurements. The numerical simulation of the master equation (Eq.~(14) in the main article) is performed by a Runge-Kutta method considering  a Fock space of up to 6 Fock states in the resonator, sufficient for the relative coupling strength $g/\omega_r=0.1$. The operators $\hat{U}$ and $\hat{S}$ are numerically built by means of projecting them into the eigenstates of the qubit-resonator Hamiltonian, $\hat{H}_0|\psi_j\rangle=\hbar\omega_j|\psi_j\rangle$, where 
\begin{equation}
\hat{H}_0=\frac{\hbar\omega_q}{2}\hat{\sigma}_z+\hbar\omega_r \hat{a}^{\dag}\hat{a} + \hbar g(\cos\theta_q\hat{\sigma}_z-\sin\theta_q\hat{\sigma}_x)(\hat{a}+\hat{a}^{\dag}).
\end{equation}
For instance, the operator $\hat{U}_k$ can be represented as
\begin{eqnarray}
\hat{U}_k&\to&\sum_{\ell,\ell'}\langle \ell|\hat{U}_k|\ell'\rangle|\ell\rangle\langle \ell'|\nonumber\\
&=&\sum_{\ell,\ell'}\int_{0}^{\infty}d\tau\nu_k(\tau)e^{i(\omega_{\ell'}-\omega_{\ell})\tau}\langle \ell|\hat{S}_k|\ell'\rangle|\ell\rangle\langle \ell'|.
\end{eqnarray}
Notice that taking into account Eq.~(16) from the main article in the above expression will lead to integrals of the form
\begin{equation}
\int_0^{\infty}d\tau e^{i(\omega_{\ell'\ell}\pm\omega)\tau}=\pi\delta(\omega_{\ell'\ell}\pm\omega)+iP\bigg(\frac{1}{\omega_{\ell'\ell}\pm\omega}\bigg),
\end{equation}
where $\omega_{\ell'\ell}=\omega_{\ell'}-\omega_{\ell}$. The second term in the above equation stands for the Cauchy principal value. The latter leads to small Lamb shifts that we have neglected in our numerical simulation. 
\begin{figure}[!hbt]
\centering
\includegraphics{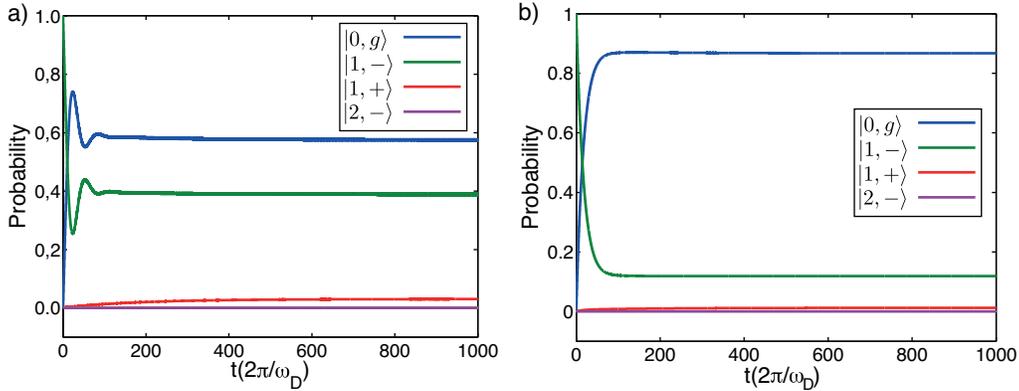}
\caption{Numerical simulation of the TCPOM master equation. The driving frequency $\omega_D$ is resonant with the transition $|0,g\rangle\to|1,-\rangle$. Blue, green, red, cyan refer to probability to populate ground $|0,g\rangle$, first $|1,-\rangle$, second $|1,+\rangle$ and third $|2,-\rangle$ eigenstates of the system. The parameters used were $\rho(0)=|1,-\rangle\langle1,-|$, $T=90$~mK, $\Gamma_r/2\pi=0.1$~MHz, $\Gamma_{\mathrm{qb}}/2\pi=15$~MHz, $\omega_D/2\pi = 4.12$~GHz, $A_{\mathrm{qb}}=A_r=90$~MHz (a), $A_{\mathrm{qb}}=A_r=12$~MHz (b). The parameters in (b) are the same as those used in Fig.~2(b) in the main article to obtain the spectroscopy traces to reproduce the experimental data.}
\end{figure}

An example of the dynamical evolution governed by the master equation can be seen in Fig.~7. The plots show the probability to populate the lowest four energy levels from Fig.~1 in the main text as function of applied driving time. The parameters used to obtain each figure are $\rho(0)=|1,-\rangle\langle1,-|$, $T=90$~mK, $\Gamma_r/2\pi=0.1$~MHz, $\Gamma_{\mathrm{qb}}/2\pi=15$~MHz, $\omega_D/2\pi=4.12$~GHz, $A_{\mathrm{qb}}=A_r=90$~MHz (12~MHz) Fig.~7(a) (Fig.~7(b)). The value of the driving frequency $\omega_D$ is resonant with the transition $|0,g\rangle\to|1,-\rangle$. Figure~7(a) features small oscillations due to a driving amplitude stronger than the qubit decay rate. Fig.~7(b) shows a smooth behavior given that the dissipative mechanisms surpass the driving microwave fields. 

In order to compare the numerical simulations with the actual data we need to take into account that the DC-SQUID detector is sensitive to the flux generated by the qubit, $\Phi_{\mathrm{qb}} = M_{\mathrm{SQ-qb}}\langle I_{\mathrm{circ}}\rangle$, where $M_{\mathrm{SQ-qb}}$ is the SQUID-qubit mutual inductance and $\langle I_{\mathrm{circ}}\rangle$ the expectation value of the circulating current operator of the qubit. The current operator has the form $\hat{I}_{\mathrm{circ}} = \alpha I_C\sin(\hat{\varphi}_{\alpha})$ with $\hat{\varphi}_{\alpha}$ the phase operator across the $\alpha$ junction. Using the two-level approximation of the flux qubit, $\langle \hat{I}_{\mathrm{circ}}\rangle= I_p\langle\hat{\sigma}_z\rangle$ calculated in the energy basis of the qubit, with $I_p$ being the persistent current. In this case it is easy to show that for a pure state $\langle\sigma_z\rangle = \pm\cos\theta_q$, with $\theta_q$ defined in the previous section. The sign depends on whether the qubit is in the ground or the excited state, respectively. For a mixed state $\hat{\rho} = p_0|0\rangle\langle0|+p_1|1\rangle\langle1|$, $\langle \sigma_z\rangle= \mathrm{Tr}(\hat{\rho}\hat{\sigma}_z)=\cos\theta_q(p_1 - p_0)$. 

\subsection*{Measurement protocol}

To perform measurements at the symmetry point we use a protocol developed in previous experiments \cite{arkady}. The qubit is initiated in an external bias flux $\Phi_i =-2.4~$m$\Phi_0$ where it produces a net magnetic field. Then a flux square pulse is applied to the qubit via the microwave line, followed by a long microwave burst. The length of the burst is sufficient to bring the system in steady-state for all resonances. At the end of the burst the flux pulse is ramped down until the qubit is brought back to the initial point. The rise time of the pulse ($\sim2~$ns) is slow enough that the Landau-Zener tunneling probability is below 1\% when the qubit is brought across the avoided-level crossing with the resonator, so the qubit population is transferred adiabatically from the symmetry point to the readout point. 

\subsection*{Selection rule for longitudinal driving}
A longitudinal single-qubit driving characterized by a $\sigma_{z}$ operator will not break the parity symmetry of the quantum Rabi model (QRM) at the qubit symmetry point. Correspondingly, the driving may induce transitions between states of the same parity subspace. For instance, transitions will be allowed between states $|\widetilde{n,-}\rangle$ to $|\widetilde{n,+}\rangle$ or to $|\widetilde{n\pm 2,\pm}\rangle$. Such transitions are activated by a proper choice of the resonance condition, with the signal amplitude being much smaller than the energy differences between higher-level states in the QRM. This assures the validity of the rotating-wave approximation. Away from the symmetry point, the QRM breaks its parity symmetry and consequently its selection rule associated with cavity and/or qubit driving. 

More explicitly, let us consider the QRM with a flux qubit at its symmetry point driven by a microwave field along the longitudinal axis $\sigma_z$. The total Hamiltonian reads:
\begin{align}
H = H_{\rm QRM} + \hbar A_{\rm z}\cos{(\omega_Dt)}\sigma_z,
\label{H_longitudinal}
\end{align}  
where $H_{\rm QRM} = \frac{\hbar\omega_q}{2}\hat{\sigma}_z+\hbar\omega_r \hat{a}^{\dag}\hat{a} + \hbar g\hat{\sigma}_x(\hat{a}+\hat{a}^{\dag})$.
The Hamiltonian (34) can be written in terms of the eigenstates of $H_{\rm QRM}$, that is, $H_{\rm QRM}|j\rangle=\hbar\omega_j|j\rangle$. Notice that the basis $\{|j\rangle\}_{j=1,\hdots,\infty}$ contain states that belong to the parity subspace $+1$ and states that belong to the parity subspace $-1$. The Hamiltonian (34) can be rewritten as
\begin{align}
H = \hbar\sum_j\omega_j|j\rangle\langle j|+\hbar A_{\rm z}\cos(\omega_Dt)\sum_{j,k}Z_{jk}|j\rangle\langle k|,
\label{H_longQRM}
\end{align} 
where $Z_{jk}=\langle j|\hat{\sigma}_z|k\rangle$. As previously mentioned, the longitudinal driving acting upon the qubit does not break the parity symmetry characterized by the parity operator $\hat{P}=e^{i\pi(\frac{1}{2}(\hat{\sigma}_{z}+\mathbb{I})+\hat{a}^{\dag}\hat{a})}$, as clearly $\sigma_z$ commutes with $\hat{P}$. The consequence is that the matrix elements $Z_{jk}$ are different from zero only for $j=k$ and for states that belong to the same parity subspace. 

If we go to an interaction representation with respect to $H_{\rm QRM}$, the Hamiltonian (35) reads
\begin{align}
H_{\rm int}(t)=\hbar A_{\rm z}\cos(\omega_Dt)\Big[\sum_{j,k>j}(Z_{jk}e^{-i\omega_{kj}t}|j\rangle\langle k|+Z_{kj}e^{i\omega_{kj}t}|k\rangle\langle j|) + \sum_{j}Z_{jj}|j\rangle\langle j|\Big],
\end{align}
where we explicitly assume that for $k>j$, $\omega_k>\omega_j$. From the above Hamiltonian it is clear that if we satisfy the condition $A_{\rm z}Z_{jj}\ll\omega_{D}$, then time-dependent corrections to the QRM eigenstates might be neglected. Moreover, if the condition $A_{\rm z}Z_{jk}\ll |\omega_{kj}+\omega_D|$ is satisfied, so that the rotating-wave approximation holds, it might be possible to activate transitions between states belonging to the same parity subspace under the resonance condition $\omega_D=\omega_{kj}$.

\section*{Acknowledgements}

We would like to acknowledge J\"urgen Lisenfeld for his contributions to the measurements and Thomas Picot and Johannes Fink for fruitful discussions. P.~F.-D. acknowledges funding from NSERC, Canada Foundation for Innovation, Ontario Ministry of Research and Innovation and Industry Canada. E.~S. would like to acknowledge funding from Spanish MINECO grants FIS2012-36673-C03-02 and FIS2015-69983-P, Basque Government grant IT472-10, UPV/EHU grant UFI 11/55, PROMISCE and SCALEQIT EU projects. G.~R. acknowledges the support from the Fondo Nacional de Desarrollo Cient\'ifico y Tecnol\'ogico (FONDECYT, Chile) under grant 1150653.

\section*{Author contributions statement}

P.~F.-D., C.~J.~P.~M.~H. and J.~E.~M. designed the experiment. P.~F.-D. fabricated the device and carried out the measurements. P.~F.-D., G.~R. and E.~S. analyzed the results theoretically. G.~R. developed the numerical analysis. P.~F.-D. wrote the manuscript. All authors reviewed the manuscript. 

\section*{Additional information}

\subsection*{Competing financial interests} The authors declare no competing financial interests.

\end{document}